# Off-Resonance Enhancement of the Refractive Index via Indirect Optical Transitions

*E.I. Battalova[1], A.I. Minibaev[1], A.V. Petrov[1] and S.S. Kharintsev[1*]*

*[1]Department of Optics and Nanophotonics, Institute of Physics, Kazan Federal University, Kremlevskaya str., 16a, Kazan, 420008, Russia*

*skharint@gmail.com

ABSTRACT

At resonance, the optical response of a homogeneous medium to incident light is governed by temporal dispersion, which induces changes in the refractive index (RI). However, recent experimental results reveal RI enhancements exceeding the fundamental limit of 5 in spatially confined systems. Herein, we study the off-resonance enhancement of RI in sub-5 nm gold nanoparticles using electronic light scattering (ELS). We discuss that the RI driven by the distribution and dynamics of excited-state charge-carriers can be locally boosted through the breaking of spatial symmetry in matter. These findings are critical for optically probing RI variations in spatially confined media using ELS.

Keywords: Electronic light scattering, refractive index, indirect transitions, electronic polarization, spatial confinement, spatially confined media, electron-photon interaction, gold nanoparticles

A visible photon entering a medium perturbs its electron system, whose response inevitably modifies the conditions for light propagation. The strength of this interaction depends on the chemical composition of the material and is driven by only two optical constants – the real refractive index (RRI) $n$ and extinction coefficient $\kappa$.[1] Beyond resonance, the material behaves as an optically transparent ($\kappa \approx 0$) medium in which the direction of propagation of a photon with momentum $k_0$ can be altered due to the expansion of momentum as strong as $nk_0$.[2]

Under resonant conditions, the electron system is driven out of equilibrium due to light absorption, resulting in significant changes in both constants.[3] A general strategy for changing RRI is based on temporal dispersion that governs the phase delay of propagating electromagnetic waves.[1,4] This magnitude, further designated as $n_t$, is basically encountered in ray optics for homogeneous bulk materials. Within the visible range, the temporal RRI is fundamentally limited to[3,4]

$$n_t(\omega) \leq \left(\frac{\omega_p^2}{\omega}\frac{dn_t}{d\omega}\right)^{1/3}, \tag{1}$$

where $\omega_p^2 = Ne^2/\varepsilon_0 m_e$ is a plasma frequency ($N$ is the concentration of polarizable electrons, $e$ and $m_e$ are charge and mass of an electron, $\varepsilon_0$ is the permittivity of vacuum), $\omega$ is a frequency of incident light. Physically, this constraint is caused by small transition dipole moments or a strong spatial localization of charge density $\rho$. At resonance, a fraction of the volume occupied by the electron cloud does not exceed 2.[1,3] However, numerous experimental measurements of the RRI of spatially dispersive media refute Eq. (1), showing that the RRI can vary in a wide range,[5–7] reaching several orders of magnitude.[8] The primary explanation for a giant broadband RRI has been the enhancement of the photonic density of states arising either from plasmonic resonances in metals[5,9], or defect-enhanced local electric fields in semiconductors.[10] A similar line of reasoning has been employed to understand the unnaturally high values $n_t$ in colloidal superlattices,[11] metafluids[12] and colloidal metamaterials.[13]

This *Letter* demonstrates that spatial confinement gives rise to a giant RRI far from resonance, showing a previously unknown enhancement mechanism. This effect is investigated using isolated and clustered sub-5 nm gold nanoparticles (Au NPs) deposited on a mica substrate. Under continuous-wave (cw) illumination, these model systems exhibit broadband background emission known as *electronic light scattering* (ELS),[14] which arises from indirect optical transitions mediated by spatial confinement. It is demonstrated that ELS spectroscopy can be employed to quantitatively probe the RRI of spatially confined systems and to reconstruct their energy band structure.

Spatial confinement plays a vital role in enhancing light-matter interactions. The smaller the scatterer, the lower-energy photon emits. This implies that nanodispersed systems scatter red rays more efficiently than blue ones – in other words – nanoscale water droplets in the atmosphere could invert the rainbow dispersion pattern. This effect is enhanced when structural moieties aggregate to form spatial percolating networks. It is essential to emphasize that, in this context, *spatial dispersion* is intrinsically tied to spatial confinement (or *spatial locality,* for instance, a phase boundary), which enables new pathways for light-matter interactions, such as *indirect optical transitions*.[15] This perspective contrasts with the conventional interpretation prevalent in the scientific community, where spatial dispersion is typically regarded as a signature of *spatial nonlocality*, described by the complex permittivity $\varepsilon(\omega, \boldsymbol{k})$ ($\boldsymbol{k}$ is the photon wavevector).[16] Instead, we introduce a spatial profile of the RRI, $n_s(\boldsymbol{r})$, that was previously coined by E. Wolf et al for deducing the strength of the scattering potential, given by $F(\boldsymbol{r}) = -k[n_s^2(\boldsymbol{r}) - 1]$.[17] The spatial RRI is intrinsically driven by charge density $\rho = -\nabla \cdot \boldsymbol{P}$ (where $\boldsymbol{P}$ is the electronic polarization) at scales $r \ll \lambda$ and asymptotically tends to its phase-delayed counterpart $n_t$.[18] The advent of a defect, whether a vacancy, inclusion, phase boundary, etc., in homogeneous solids locally breaks spatial symmetry, which allows an incident photon to induce a polarization current, boosting the spatial RRI, $n_s$. The optically induced polarization current transfers, like a phonon, additional momentum to an electron, and thereby enabling indirect transitions. Fig. 1 shows a light scattering mechanism in spatially confined and bulk gold. Experimental evidence of this physical phenomenon is provided by broadband background emission observed in confined metallic structures – such as rough films, sharp tips, atomic cavities and protrusions – commonly referred to as ELS.[19–25] This effect closely resembles electronic Raman scattering,[26,27] in which the initial and final electronic states differ, and an electron moves between them without a change in momentum. On the other hand, this is reminiscent of the Compton effect, in which the electron's momentum can be altered during optical transition. The underlying physics stems from the increased uncertainty in the momentum of the optically induced polarization current, generating a quasi-stationary electrical near-field ($E(r) \sim 1/r^3$),[16,28] sometimes referred to as *a near-field photon*,[29] and the slope of the *sp*-conduction band.[21]

Light absorption in spatially confined structures can also be enhanced by indirect transitions;[30,31] however, the efficiency of this process is limited by the absorbed photon energy and the associated momentum uncertainty. This mechanism becomes predominant for heterogeneous media containing a wide diversity of sizes, thereby squeezing momentum uncertainty. It is important to stress that the standard relationship $A_t = 1 - R - T$ (where $R$ and $T$ is reflection and transmission) for quantifying absorption by heterogeneous media at a given frequency should be applied with caution since the total losses $A_t$ involve not only direct/indirect absorption but also direct/indirect scattering, all of which are inherently convoluted and not easily separable.[14] The total absorption can be estimated by the following generalized expression $A(\omega) = 1 - \int_0^{\omega}[R_\omega(\omega') + T_\omega(\omega')]d\omega'$ (where $R_\omega$ and $T_\omega$ are spectral densities of $R$ and $T$, respectively), which considers the losses due to indirect light scattering processes.

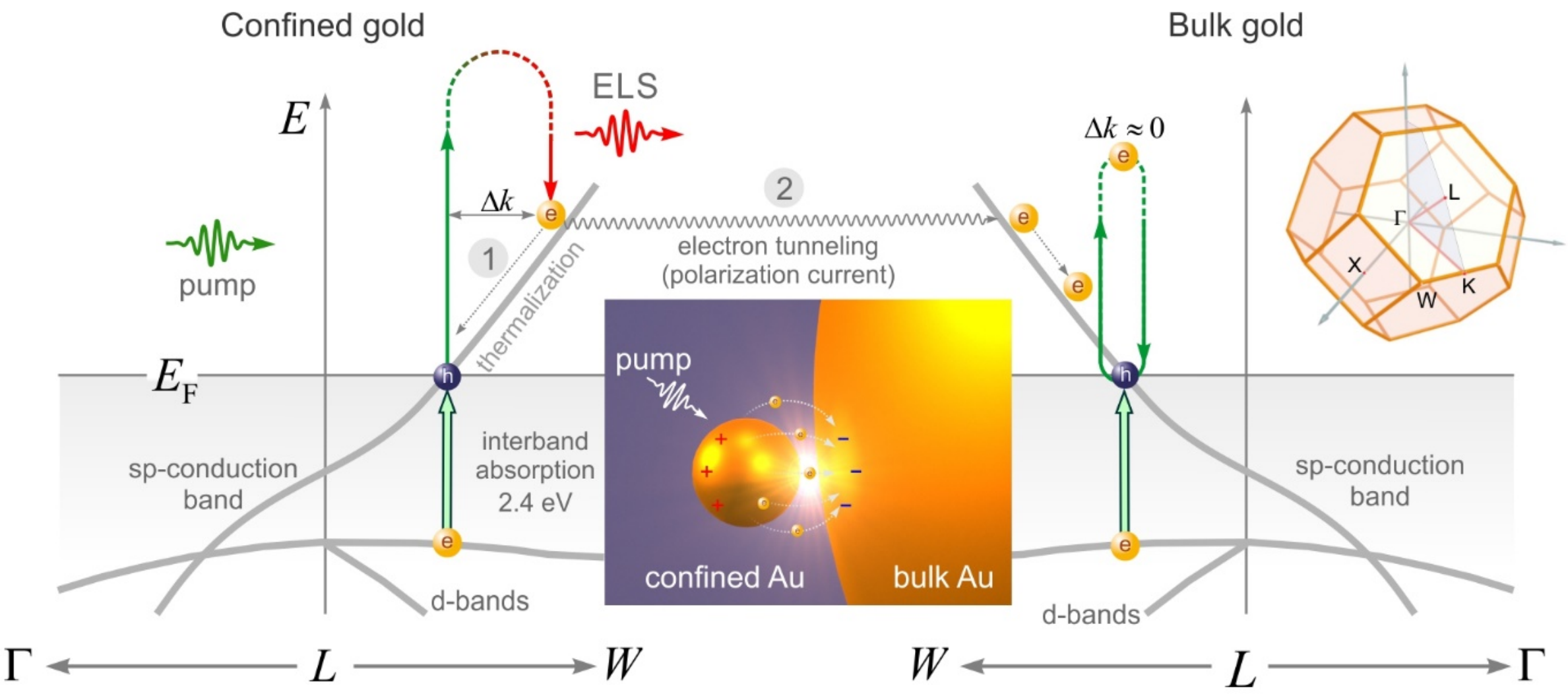

**Figure 1.** Schematic illustration of interband absorption and electronic light scattering in both confined (left panel) and bulk gold (right panel). The central inset shows a tunneling mechanism for excited electrons from confined to bulk gold. The right-up inset displays the Brillouin zone for gold.

An incident photon, unable to transfer its momentum to an electron because of electron-photon momentum mismatching, induces the polarization current within a spatially confined structure. This electronic excitation, with momentum enhanced by spatial confinement, enables indirect transitions (Fig. 1). It is precisely the medium rather than the electromagnetic field, which transfers additional momentum to the electron. Indeed, the cycle-averaged optical force exerted on the polarizable electron along the $z$ axis is given by $\langle \boldsymbol{F} \rangle = \alpha'' k \langle \boldsymbol{E}^2 \rangle \nabla[n_s(\boldsymbol{r})\hat{\boldsymbol{e}}_z \boldsymbol{r}]$ (where $\alpha''$ are the imaginary part of atomic polarizability, $\hat{\mathbf{e}}_z$ is the unit vector showing the direction of the incident electromagnetic wave, $\boldsymbol{E}$ is the electric field). This force is enhanced by two to three orders of magnitude when $r \ll \lambda$, owing to the large spatial gradient of the local RRI introduced by the spatially confined medium. In the far-field when $r > \lambda$ and $n_s(\boldsymbol{r}) \to n_t(\omega)$ this force transforms to $\langle \boldsymbol{F} \rangle = \alpha'' \omega \langle \boldsymbol{E} \times \boldsymbol{B} \rangle$ (here $\boldsymbol{B}$ is the magnetic field), which is proportional to the average field momentum – normally insufficient to be transferred to an electron.[16] Under the influence of this force, the Fermi electron is excited to higher-energy states within the *sp*-conduction band and subsequently thermalizes back (this process is labelled as '1'), releasing heat. This process is accompanied by the scattering of a redshifted photon, which corresponds to ELS.[25]

Under resonance, interband absorption at 2.4 eV (near the $L$ point of the Brillouin zone) additionally populates the Fermi level (Fig. 1). The balance between interband absorption and indirect transitions induces charge density oscillations within spatially confined metals, known as localized plasmon resonance. The $Q$-factor of this resonance is driven by both Landau damping[32,33] and Lamb mode excitation.[34,35] In fact, the indirect transitions are directly linked to Landau damping. Specifically, slow electrons in high-energy states, whose kinetic energy is approximately $K \approx 1/2\mathcal{L} \cdot dE/dk$ ($\mathcal{L}$ is a spatial confinement, $E(k)$ is the electron energy in the conduction band (details in Section I, the Supplementary Information (SI))), can be captured by an oscillating charge-carrier cloud. This cloud becomes damped due to collision-free energy transfer to the slow electrons with $dE/dk \approx 0$. At the nanogap (the central inset in Fig. 1), hot electrons with the maximal slope, $dE/dk$, can tunnel from the Au NP to the bulk Au, generating a giant polarization current within it (this process is labelled as '2' in Fig. 1). Clearly, this effect is significantly enhanced in the case of a dimer consisting of identical Au spheres. It leads to a redistribution of the charge density, thereby enabling a giant enhancement of the RRI. Physics behind the enhancement is attributed to – at least in principle – indirect transitions which must be taken into account when calculating the RRI:[14]

$$n_s^2(\boldsymbol{r}) = 1 + \frac{e^2}{\pi^2 m_e} \sum_{cv} \int_{BZ} \frac{f_{cv}^{\varkappa}(\boldsymbol{k})}{\Omega_{cv}^2(\boldsymbol{k}) - \omega^2} d\boldsymbol{k} , \tag{2}$$

where $\Omega_{cv}$ is a vibronic frequency corresponding to optical transitions between the electronic states $|v\rangle$ and $|c\rangle$, $\boldsymbol{\varkappa}$ is a light polarization direction. The oscillator strength as a function of momentum is modified as follows

$$f_{cv}^{\varkappa}(\boldsymbol{k}) = \frac{2m_e}{\hbar}|D_{cv}^{\varkappa}(\boldsymbol{k})|^2, \qquad (3)$$

here $\hbar$ is the Planck's constant, $D_{cv}^{\varkappa}(\boldsymbol{k}) = \langle c|e^{k(r)\varkappa\boldsymbol{r}}\,\partial/\partial\boldsymbol{r}\,|v\rangle$ is the transient electrical dipole moment taking spatial confinement into account. In Eq. (2), integration runs over the entire Brillouin zone, indicating that all optical transitions are allowed, including those between the states with the same symmetry. The total number of excited electrons, $N$, increases significantly and the oscillator strength sum rule is transformed to:

$$\sum_{cv}\int_{BZ} f_{cv}^{\varkappa}(\boldsymbol{k})d\boldsymbol{k} = \frac{2m_e}{\hbar}N. \qquad (4)$$

Spatial confinement results in the enhanced electronic polarization or a delocalization of the near-field. This effect is especially evident in plasmon resonance that decays rapidly in metallic nanoparticles smaller than 5 nm, because indirect scattering and interband absorption become imbalanced for such nanoparticles.

Near-field delocalization can be enhanced by clustering tiny metallic nanoparticles, forming extended percolating networks.[36] To validate this hypothesis, we consider a model system consisting of two spherical Au NPs, as illustrated in the top-left inset of Fig. 2. Our goal is to calculate the electric field strength $E_x$ in the vicinity of the Au NP (1) as a function of the distance $d$ between two nanoparticles. The $E_x$-field enhancement is calculated at a point located

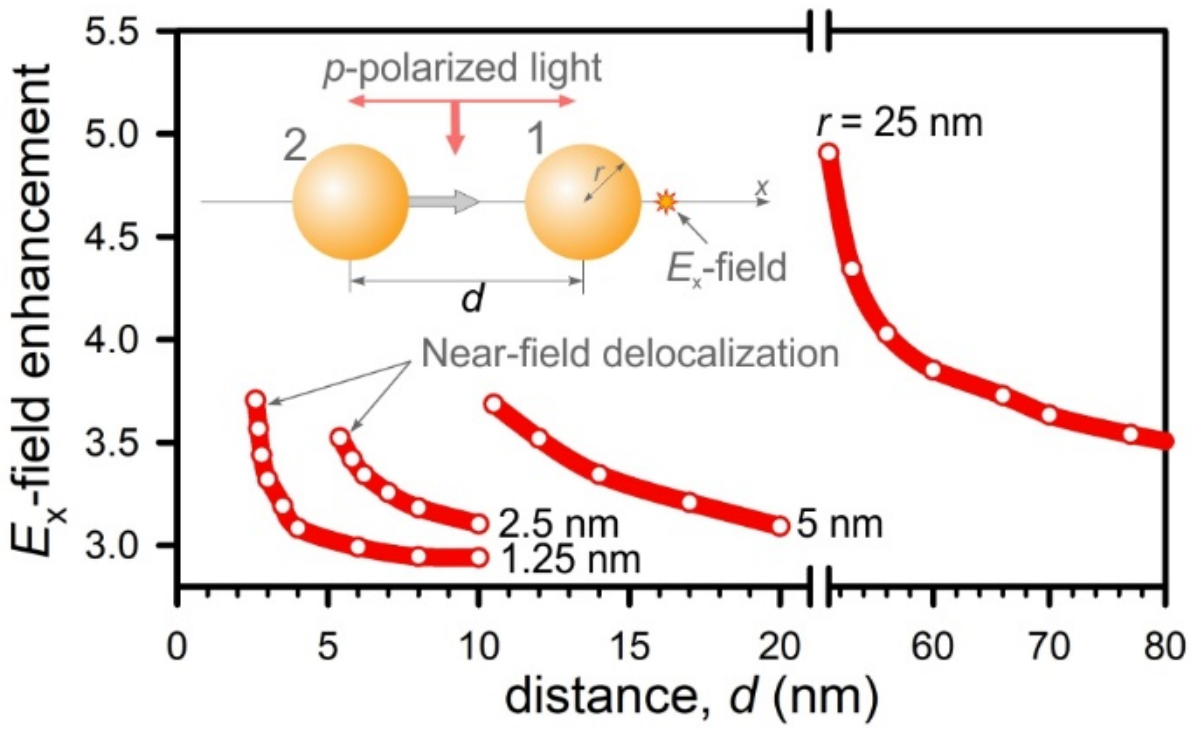

**Figure 2.** FDTD simulation of the electric $E_x$-field enhancement factor at the point marked with a star on the $x$-axis when two similar Au spheres labelled as '1' and '2' with varying radii are separated by a distance $d$ (633 nm excitation).

0.5 nm on the right from the surface of Au NP (1) along the $x$-axis. It is important to stress that we calculate the electric field at the point in which spatial inhomogeneity holds intact. The charge-carrier density in Au NP (1) significantly enhances due to the excitation of a gap-plasmon between Au NP (1) and Au NP (2), which is dependent on $d$. This dimer is illuminated by a $p$-polarized cw 633 nm laser light. FDTD simulations (ANSYS Solver) were conducted for nanoparticles with radii of 1.25, 2.5, 5 and 25 nm, as shown in Fig. 2 (Section II, the SI). All isolated nanoparticles exhibit a threefold increase in $E_x$ independent of plasmonic resonance, except for the 25 nm Au NP, which shows a slightly higher enhancement of 3.6 due to the extended plasmon resonance tail. We clearly see that as the nanoparticles approach each other, the electric field increases, and a stronger $E_x$-field gradient is achieved for smaller nanoparticles. Thus, we infer that clustered metallic nanoparticles can enhance the polarization current and delocalize the near-field, thereby further increasing the RRI. The enhancement, in this context, occurs due to the gap plasmon excitation between the Au NPs. This theoretical result explains the increased charge density $\rho$ at sites with broken spatial symmetry. This is the reason why the normal component of the electrical field is discontinuous at the phase boundary.

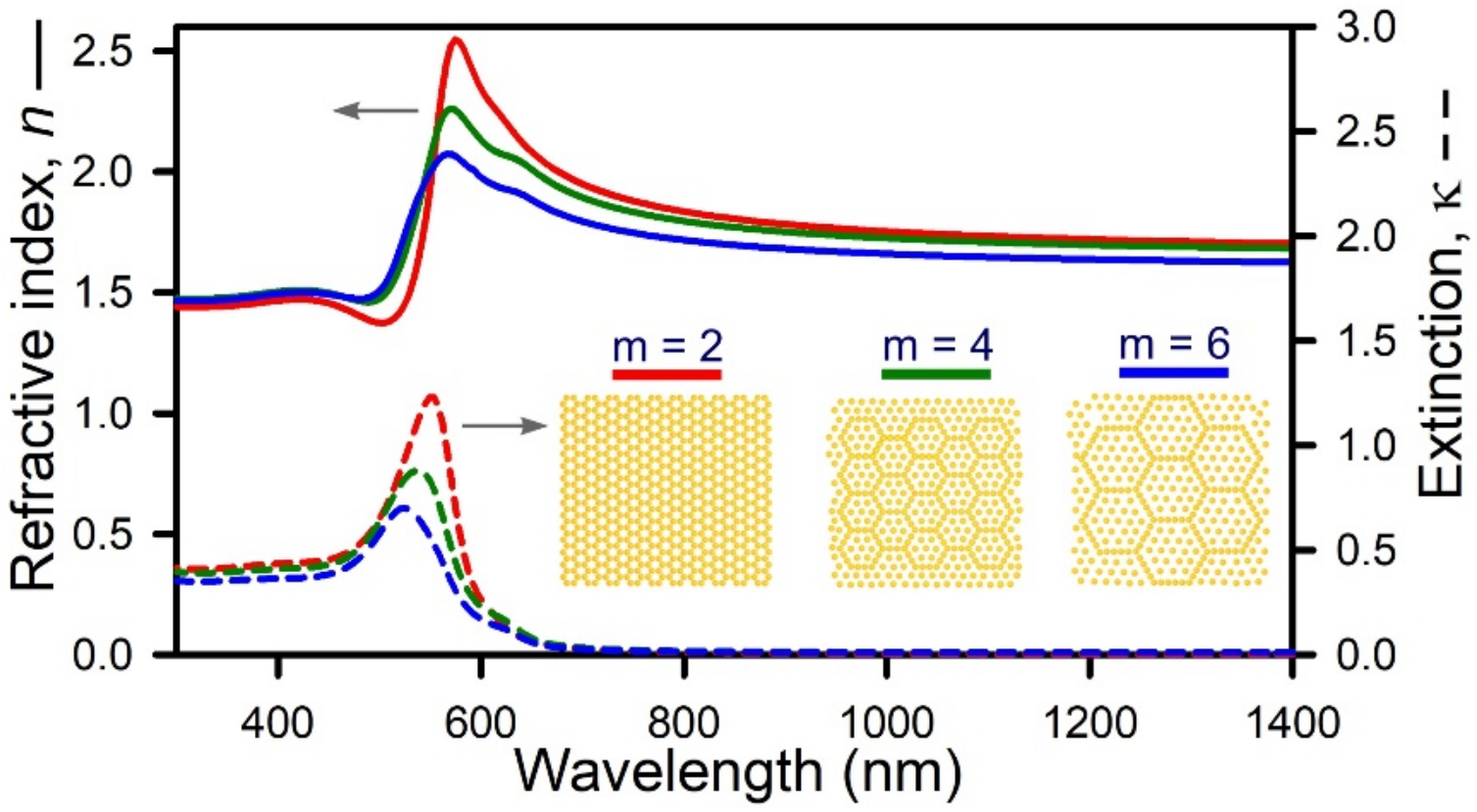

**Figure 3.** Numerical calculation of the real refractive index and extinction coefficient of clustered 5 nm Au NPs.

Defect-rich solids behave as high-index materials, compared to their homogeneous counterparts. One example, the ferroelectric perovskite KTN:Li, which has the RRI of 2.5 at room temperature and exhibits its increase up to 26 during the phase transition, where structural fluctuations are huge.[6] Seungwoo Lee and co-workers showed that clusters of 60 nm Au polyhedra ( $n_s = 10.2$ ) surpass their spherical counterparts ( $n_s = 5.3$ ) due to spatial confinement ensuring stronger polarization currents.[5]

FDTD simulations of the refractive index and extinction for clustered 5 nm Au NPs, conducted using the $S$-parameter retrieval method,[37] are shown in Fig. 3. For simplicity, we employed a two-dimensional hexagonal arrangement with varying numbers of Au spheres per hexagon segment ($m = 2, 4, 6$) and a fixed interparticle distance of 0.5 nm. To maintain a constant filling factor, additional Au NPs are randomly dispersed throughout the structure, with a minimum interparticle distance of 1 nm. The most densely clustered Au NPs ($m = 2$) shows a maximum refractive index of 1.7 off-resonance and 2.5 at resonance. Upon diluting the cluster, the resonant RRI decreases to 2.2 ($m = 4$) and 2.0 ($m = 6$), while the off-resonance values change slightly. The $Q$-factor of the plasmon resonance drops and the extinction coefficient reveals a blueshift of 30 nm. This trend persists until Au NPs become isolated and exhibit localized plasmon resonances. In contrast, when all Au NPs merge to form a homogeneous bulk Au film the broadband RRI drops to 0.2-0.3 in the long-wavelength region (Section III, the SI). Thus, spatial confinement serves as the underlying mechanism responsible for the observed giant broadband RRI.

To experimentally validate the results of our numerical analysis, we have conducted ELS experiments on mica surfaces coated with 5 nm Au NPs. Two types of coatings were applied: 1) 2-ammonioethyl di-tert-butylphosphonium (ADTB)-coated isolated Au NPs were randomly distributed on the mica surface; 2) bis(p-sulfonatophenyl)phenylphosphine (BSPP)-coated Au NPs were self-clustered on the mica surface, forming an extended percolating film (details in Section IV, the SI). The interparticle distance, controlled by the ligand size, varies within the range of 1-2 nm.

Fig. 4a shows atomic force microscopy (AFM) images of isolated (blue) and clustered (red) 5 nm Au NPs. Cross-section profiles along the white dashed lines for both cases are presented in Fig. 4b. Some of isolated Au NPs stick to form agglomerates which distort the optical response due to the large diversity of their sizes. In contrast, clustered Au NPs shape an infinitely extended percolating monolayer film, whose optical response is highly stable and reproducible. Fig. 4c shows low-energy and high-energy ELS spectra of isolated and clustered 5 nm Au NPs, measured with a confocal spectrometer NTEGRA SPECTRA (NT-MDT Co.). A low-energy central continuum is attributed to optical transitions near the Fermi level (Fig. 1). The isolated 5 nm Au NPs exhibit spheroidal (1,0) and quadripolar (1,2) elastic Lamb modes[38,39] at 22 $cm^{-1}$ and 12 $cm^{-1}$ which are superimposed on the ELS continuum. The lack of the quadripolar mode narrows the central peak for clustered Au NPs, as seen in Fig. 4c. On the other hand, this attenuated mode, which influences the effective size of the Au NPs, increases momentum uncertainty and promotes intense indirect transitions. The inhomogeneous broadening of the low ELS band from isolated Au NPs is enhanced due to the formation of larger agglomerates, which reduce spatial localization and relax momentum uncertainty. These

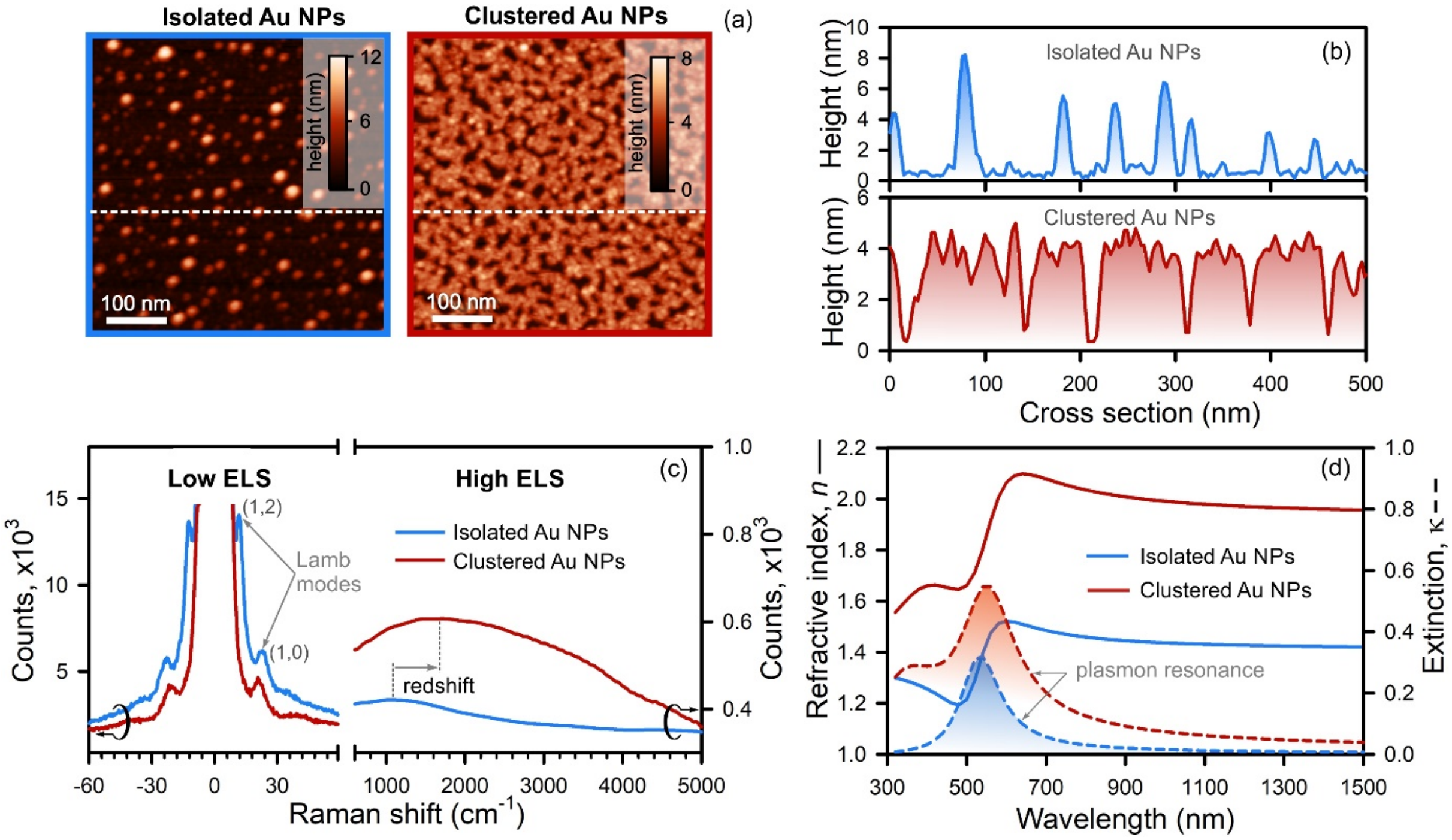

**Figure 4.** An analysis of isolated (blue) and clustered (red) 5 nm Au NPs. (a) AFM images and (b) cross-sections along white dashed lines. (c) Low-energy and high-energy ELS. (d) The measured refractive index (solid curve) and extinction coefficient (dashed curve) vs wavelength.

factors also weaken the intensity of high-energy ELS spectra. Clustering Au NPs results in a redshift of the high-energy ELS continuum. A similar behavior is observed for clustered Au

NPs with a radius of 2.5 nm (Section V, the SI). However, this effect vanishes for 1.5 nm Au NPs (Section V, the SI) because structural differences between isolated and clustered nanostructures become negligible. This intriguing observation hints at increasing momentum uncertainty, which reinforces the oscillator strength and, as a consequence, the RRI, Eq. (2). This assumption is supported by direct measurements of RRI and extinction coefficient using a spectroscopic ellipsometer VASE (Woollam Co., Inc.) (Fig. 4d). We observe a broadband enhancement of the RRI for clustered Au NPs within the visible and infrared region. The plasmon resonance exhibits a redshift of 20 nm upon clustering Au NPs. The experimental results are close to those obtained by FDTD simulation (Fig. 3). Indeed, the excess "absorption" arising from ELS contributes to an increase in the off-resonance RRI, thereby supporting the validity of our model (Eq. 2). The 20% overestimation of the experimental values arises from indirect transitions, which are not included in the calculations but inevitably contribute to the charge density in the conduction band. The nexus between RRI and ELS is established through the dynamics of charge density $\rho$ – ELS enhances the excited-state electron population, whereas RRI serves as a sensitive probe of this charge density redistribution. The ELS intensity is determined by the $k$-dependent photonic density of states $\mathcal{F}(k)$ and the population of charge carriers, driven by Fermi-Dirac statistics $f_{FD}(k)$:[14]

$$I_{ELS}(k) = C\int \mathcal{F}(k-k')\mathcal{E}(k')f_{FD}(k')dk', \tag{5}$$

where $C$ is a constant proportional to the scattering cross-section, $\mathcal{E}(k) = \hbar\omega_0 - E(k)$ is the scattered photon energy, $E(k)$ is the electron energy in the conduction band. It is important to emphasize that ELS is a pre-relaxation physical process distinct from both photoluminescence and Raman scattering. This phenomenon is predominantly observed in defect-rich materials, including liquids. ELS depends on the excitation wavelength, but its spectral shape is primarily determined by the slope of the energy band landscape. In the case of flat energy band valleys, ELS vanishes.

In 1981, Yablonovitch demonstrated that the incident light intensity is enhanced by a factor of $n_s^2$ in ergodic media, because $\mathcal{F}(k) \sim n_s^2$, where $n_s$ is a local refractive index.[40] The same reasoning is valid for light scattering intensity, $\mathcal{E}(k) \sim n_s^2$ . Based on these findings, we infer that the ELS intensity scales as $I_{ELS}(k) \sim n_s^4$, a dependence that disappears in homogeneous media characterized by $n_t$. Like vibrational Raman scattering, the ELS intensity is proportional to a factor $[\omega \pm \Omega_{cv}(k)]^4$ and, therefore, RRI can be directly linked to a Raman shift: $n_s \sim [\omega \pm \Omega_{cv}(k)]$. This statement is verified by experimental observations of both ELS and RRI (Fig. 4c,d). The physical mechanism underlying RRI enables a direct connection between this parameter and the field enhancement factor $g = |\boldsymbol{E}|/|\boldsymbol{E}_0|$ (where $\boldsymbol{E}_0$ and $\boldsymbol{E}$ are the incident and enhanced electrical fields) that shows how the incident electrical field is amplified near a spatially confined structure. Ultimately, both parameters are governed by the distribution and dynamics of excited-state charge-carriers. The field enhancement factor $g$ is widely used to quantitatively estimate the intensities of molecular peaks in TERS/SERS experiments,[41–44] $I_{TERS}(\omega) \sim g^4$,[45–48] whereas the spatial refractive index $n_s$ serves as a reliable indicator of broadband background emission, $I_{ELS}(k) \sim n_s^4$, that is an integral part of TERS/SERS spectra. Both phenomena are fundamentally interlinked through a common underlying mechanism. The polarization current (the process '2' in Fig. (1)) contributes to the enhancement of molecular vibrations or phonons, whereas electronic light scattering is seen as a side background emission, showing the strength of indirect transitions.

In conclusion, we emphasize that the off-resonance enhancement of RRI originates from indirect transitions mediated by spatial confinement. The additional momentum transferred to the electron is generated by the spatially confined medium itself rather than a confined photon, as previously thought. Our approach explains not only the increased refractive index, but also anomalous photoheating,[14] amplified electrical conductivity[49] and low-intensity pump nonlinear effects.[49] A distinctive feature of spatial refractive index is electronic light scattering which can be used for optically probing the RRI of spatially confined media. In contrast, optical spectrophotometry and spectroscopic ellipsometry are primarily suited for homogeneous bulk materials. Yet, ELS spectroscopy can be employed to reconstruct the energy band structure. These findings are critical for the development of high-index conductive transparent materials for thermo-optical and photo-electrical applications in optoelectronics and photovoltaics.

**ASSOCIATED CONTENT**

**Supporting Information**

Details on FDTD simulation of electric field enhancement and the refractive index of isolated and clustered Au NPs, sample preparation, ELS spectra of isolated and clustered 1.5 and 2.5 nm Au NPs.

## AUTHOR INFORMATION

**Corresponding Author** *E-mail: skharint@gmail.com

**ORCID**

Elina Battalova: 0009-0005-5747-1799

Aidar Minibaev: 0009-0007-6621-5820

Andrey Petrov: 0000-0003-3202-2477

Sergey Kharintsev: 0000-0002-5788-3401

**Notes**

The authors declare no competing financial interest

## ACKNOWLEDGEMENT

The authors gratefully acknowledge A.B. Kotlyar and L. Katrivas for fabricating a series of samples. We are especially thankful to Ara Apkarian and Andrew Shubin for fruitful discussion. The research was funded by the subsidy allocated to Kazan Federal University for the state assignment in the sphere of scientific activities (FZSM-2025-0004). The authors acknowledge a technical support from Ostec-ArtTool and NT-MDT.

## REFERENCES

1. Khurgin, J. B. Energy and Power Requirements for Alteration of the Refractive Index. *Laser Photonics Rev.* **2024**, *18* (4), 2300836.
2. Kharintsev, S. S.; Battalova, E. I.; Mkhitaryan, V.; Shalaev, V. M. How Near-Field Photon Momentum Drives Unusual Optical Phenomena: Opinion. *Opt. Mater. Express* **2024**, *14* (8), 2017–2022.

3. Khurgin, J. B. Expanding the Photonic Palette: Exploring High Index Materials. *ACS Photonics* **2022**, *9* (3), 743–751.
4. Shim, H.; Monticone, F.; Miller, O. D. Fundamental Limits to the Refractive Index of Transparent Optical Materials. *Adv. Mater.* **2021**, *33* (43), 2103946.
5. Kim, N.; Huh, J.-H.; Cho, Y.; Park, S. H.; Kim, H. H.; Rho, K. H.; Lee, J.; Lee, S. Achieving Optical Refractive Index of 10-Plus by Colloidal Self-Assembly. *Small* **2024**, *20* (45), 2404223.
6. Di Mei, F.; Falsi, L.; Flammini, M.; Pierangeli, D.; Di Porto, P.; Agranat, A. J.; DelRe, E. Giant Broadband Refraction in the Visible in a Ferroelectric Perovskite. *Nat. Photonics* **2018**, *12* (12), 734–738.
7. Andreoli, F.; Gullans, M. J.; High, A. A.; Browaeys, A.; Chang, D. E. Maximum Refractive Index of an Atomic Medium. *Phys. Rev. X* **2021**, *11* (11), 011026.
8. Rivera, N.; Kaminer, I.; Zhen, B.; Joannopoulos, J. D.; Soljačić, M. Shrinking Light to Allow Forbidden Transitions on the Atomic Scale. *Science* **2016**, *353* (6296), 263–269.
9. Huh, J.-K.; Lee, J.; Lee, S. Soft Plasmonic Assemblies Exhibiting Unnaturally High Refractive Index. *Nano Lett.* **2020**, *20* (7), 4768–4774.
10. Meixner, A. J.; Bopp, M. A.; Tarrach, G. Direct Measurement of Standing Evanescent Waves with a Photon-Scanning Tunneling Microscope. *Appl. Opt.* **1994**, *33* (34), 7995–8000.
11. Lee, S. Colloidal Superlattices for Unnaturally High-Index Metamaterials at Broadband Optical Frequencies. *Opt. Express* **2015**, *23* (22), 28170–28181.
12. Kim, K.; Yoo, S.; Huh, J.-H.; Park, Q.-H.; Lee, S. Limitations and Opportunities for Optical Metafluids To Achieve an Unnatural Refractive Index. *ACS Photonics* **2017**, *4* (9), 2298–2311.
13. Huh, J.-H.; Kim, K.; Im, E.; Lee, J.; Cho, Y.; Lee, S. Exploiting Colloidal Metamaterials for Achieving Unnatural Optical Refractions. *Adv. Mater.* **2020**, *32* (51), 2001806.
14. Kharintsev, S. S.; Battalova, E. I. Heat Generation in Spatially Confined Solids through Electronic Light Scattering. *Nanophotonics* **2025**, *14* (14), 2411–2418.

15. Shalaev, V. M.; Douketis, C.; Haslett, T.; Stuckless, T.; Moskovits, M. Two-Photon Electron Emission from Smooth and Rough Metal Films in the Threshold Region. *Phys. Rev. B* **1996**, *53* (16), 11193–11206.
16. Novotny, L.; Hecht, B. *Principles of Nano-Optics*; Cambridge University Press, 2012.
17. Wolf, E.; Nieto-Vesperinas, M. Analyticity of the Angular Spectrum Amplitude of Scattered Fields and Some of Its Consequences. *J. Opt. Soc. Am. A* **1985**, *2* (6), 886–890.
18. Feynman, R. P.; Leighton, R. B.; Sands, M. *The Feynman Lectures on Physics*, Vol. 1; Basic Books, 2010.
19. Pilz, W.; Kriegsmann, H. The Nature of the so Called "Raman Background". *Z. Phys. Chemie* **1987**, 2680 (1), 215–216.
20. Gass, A. N.; Kapusta, O. I.; Klimin, S. A.; Mal'shukov, A. G. The Nature of the Inelastic Background in Surface Enhanced Raman Scattering Spectra of Coldly-Deposited Silver Films. The Role of Active Sites. *Solid. State Commun.* **1989**, *71* (9), 749–753.
21. Beversluis, M.; Bouhelier, A.; Novotny, L. Continuum Generation from Single Gold Nanostructures through Near-Field Mediated Intraband Transitions. *Phys. Rev. B* **2003**, *68* (11), 115433.
22. Mahajan, S.; Cole, R. M.; Speed, J. D.; Pelfrey, S. H.; Russell, A. E.; Bartlett, P. N.; Barnett, S. M.; Baumberg, J. J. Understanding the Surface-Enhanced Raman Spectroscopy "Background". *J. Phys. Chem. C* **2010**, *114* (16), 7242–7250.
23. Hugall, J. T.; Baumberg, J. J. Demonstrating Photoluminescence from Au is Electronic Inelastic Light Scattering of a Plasmonic Metal: The Origin of SERS Backgrounds. *Nano Lett.* **2015**, *15* (4), 2600–2604.
24. Inagaki, M.; Isogai, T.; Motobayashi, K.; Lin, K.-Q.; Ren, B.; Ikeda, K. Electronic and Vibrational Surface-Enhanced Raman Scattering: From Atomically Defined Au(111) and (100) to Roughened Au. *Chem. Sci.* **2020**, *11* (36), 9807–9817.
25. Kharintsev, S. S.; Battalova, E. I.; Noskov, A. I.; Merham, J.; Potma, E. O.; Fishman, D. A. Photon-Momentum-Enabled Electronic Raman Scattering in Silicon Glass. *ACS Nano* **2024**, *18* (13), 9557–9565.
26. Cai, W. R.; Mei, J. J.; Shen, W. Z. A Proposal for Complete Interband Absorption in Indirect Semiconductors. *Phys. B* **2004**, *352* (1–4), 179–184.

27. Ismailov, T. G.; Nazanly, R. A. Interband Electronic Raman Scattering in Semiconductors Induced by Nonparabolicity. *Phys. Stat. Sol.* **1991**, *164* (2), 553–560.
28. Landau, L. D.; Lifshitz, E. M. *Course of Theoretical Physics: Electrodynamics of Continuous Media*, 2nd ed.; Pergamon Press, 1984.
29. Bharadwaj, P.; Deutsch, B; Novotny, L. Optical Antennas. *Adv. Opt. Photonics* **2009**, *1* (3), 438–483.
30. Yamaguchi, M.; Nobusada, K. Indirect Interband Transition Induced by Optical Near Fields with Large Wave Numbers. *Phys. Rev. B* **2016**, *93*, 195111.
31. Kharintsev, S. S.; Noskov, A. I.; Battalova, E. I.; Katrivas, L.; Kotlyar, A. B.; Merham, J. G.; Potma, E. O.; Apkarian, V. A.; Fishman, D. A. Photon Momentum Enabled Light Absorption in Silicon. *ACS Nano* **2024**, *18*, 26532–26540.
32. Hartland, G. V. Optical Studies of Dynamics in Noble Metal Nanostructures. *Chem. Rev.* **2011**, *111* (6), 3858–3887.
33. Hartland, G. V.; Besteiro, L. V.; Johns, P.; Govorov, A. O. What's so Hot about Electrons in Metal Nanoparticles? *ACS Energy Lett.* **2017**, *2* (7), 1641–1653.
34. Lamb, H. On the Vibrations of an Elastic Sphere. *Proc. of London Math. Soc.* **1882**, *13* (1), 189–212.
35. Su, Z.; Ye, L.; Lu, Y. Guided Lamb Waves for Identification of Damage in Composite Structures: a Review. *J. Sound Vib.* **2006**, *295* (3–5), 753–780.
36. Markel, V. A.; Shalaev, V. M.; Zhang, P.; Huynh, W.; Tay, L.; Haslett, T. L.; Moskovits, M. Near-Field Optical Spectroscopy of Individual Surface-Plasmon Modes in Colloid Clusters. *Phys. Rev. B* **1999**, *59* (16), 10903–10909.
37. Smith, D. R.; Vier, D. C.; Koschny, Th.; Soukoulis, C. M. Electromagnetic Parameter Retrieval from Inhomogeneous Metamaterials. *Phys. Rev. E* **2005**, *71* (3), 036617.
38. Bayle, M.; Combe, N.; Sangeetha, N. M.; Viau, G.; Carles, R. Vibrational and Electronic Excitations in Gold Nanocrystals. *Nanoscale* **2014**, *6* (15), 9157–9165.
39. Martinet, Q.; Berthelot, A.; Girard, A.; Donoeva, B.; Comby-Zerbino, C.; Romeo, E.; Bertorelle, F.; van der Linden, M.; Tarrat, N.; Combe, N.; Margueritat, J. Performances of the Lamb Model to Describe the Vibrations of Gold Quantum-Sized Clusters. *J. Phys. Chem. C* **2020**, *124* (35), 19324–19332.

40. Yablonovitch, E. Statistical Ray Optics. *J. Opt. Soc. Am.* **1982**, *72* (7), 899–907.
41. Zrimsek, A. B.; Chiang, N.; Mattei, M.; Zaleski, S.; McAnally, M. O.; Chapman, C. T.; Henry, A.-I.; Schatz, G. C.; Van Duyne, R. P. Single-Molecule Chemistry with Surface- and Tip-Enhanced Raman Spectroscopy. *Chem. Rev.* **2017**, *117* (11), 7583–7613.
42. Anderson, N.; Hartschuh, A.; Novotny, L. Near-Field Raman Microscopy. *Mater. Today* **2005**, *8* (5), 50–54.
43. Zhang, R.; Zhang, Y.; Dong, Z. C.; Jiang, S.; Zhang, C.; Chen, L. G.; Zhang, L.; Liao, Y.; Aizpurua, J.; Luo, Y.; Yang, J. L.; Hou, J. G. Chemical Mapping of a Single Molecule by Plasmon-Enhanced Raman Scattering. *Nature* **2013**, *498* (6), 82–86.
44. Lee, J.; Crampton, K. T.; Tallarida, N.; Apkarian, V. A. Visualizing Vibrational Normal Modes of a Single Molecule with Atomically Confined Light. *Nature* **2019**, *568* (4), 78–82.
45. Richards, D.; Milner, R. G.; Huang, F.; Festy, F. Tip-Enhanced Raman Microscopy: Practicalities and Limitations. *J. Raman Spectrosc.* **2003**, *34* (9), 663–667.
46. Hartschuh, A.; Beversluis, M. R.; Bouhelier, A.; Novotny, L. Tip-Enhanced Optical Spectroscopy. *Phil. Trans. R. Soc. A.* **362** (1817), 807–819.
47. Cançado, L. G.; Hartschuh, A.; Novotny, L. Tip-Enhanced Raman Spectroscopy of Carbon Nanotubes. *J. Raman Spectrosc.* **2009**, *40* (10), 1420–1426.
48. Schultz, Z. D.; Marr, J. M.; Wang, H. Tip Enhanced Raman Scattering: Plasmonic Enhancements for Nanoscale Chemical Analysis. *Nanophotonics* **2014**, *3* (1-2), 91–104.
49. Kharintsev, S. S.; Battalova, E. I.; Matchenya, I. A.; Nasibulin, A. G.; Marunchenko, A. A.; Pushkarev, A. P. Extreme Electron-Photon Interaction in Disordered Perovskites. *Adv. Sci.* **2025**, *12* (5), 2405709.